\documentclass[preprint,review,12pt]{elsarticle}

\let\today\relax
\makeatletter
\def\ps@pprintTitle{%
    \let\@oddhead\@empty
    \let\@evenhead\@empty
    \def\@oddfoot{\footnotesize\itshape
         { } \hfill\today}%
    \let\@evenfoot\@oddfoot
    }
\makeatother

\usepackage{graphicx}
\usepackage{amsmath}
\usepackage{amssymb}
\usepackage{xcolor}
\usepackage{epstopdf, epsfig}
\usepackage{multirow}
\usepackage{booktabs} %
\usepackage{hyperref}
\usepackage{lineno}
\usepackage{longtable} %
\usepackage[utf8]{inputenc}
\usepackage{mathtools}
\hypersetup{
colorlinks=true,
linkcolor=black,
citecolor=black,
urlcolor=blue
}

\begin{document}
\begin{frontmatter}

\author[label1]{Xiangwei Li}
\author[label2]{Dongdong Wan}
\author[label2]{Mengqi Zhang}
\author[label1]{Huanshu Tan\corref{cor1}}
\ead{tanhs@sustech.edu.cn}
\cortext[cor1]{Corresponding author}

\affiliation[label1]{organization={Multicomponent Fluids group, Center for Complex Flows and Soft Matter Research, Department of Mechanics and Aerospace Engineering, Southern University of Science and Technology},
            city={Shenzhen},
            postcode={518055},
            state={Guangdong},
            country={P.R. China}}

\affiliation[label2]{organization={Department of Mechanical Engineering, National University of Singapore},
            addressline={ 9 Engineering Drive 1},
            postcode={117575},
            country={Republic of Singapore}}

\title{Marangoni Interfacial Instability Induced by Solute Transfer Across Liquid-Liquid Interfaces}

\begin{abstract}
{
This study presents analytical and numerical investigations of Marangoni interfacial instability in a two-liquid-layer system with constant solute transfer across the interface. 
While previous research has established that both diffusivity and viscosity ratios affect hydrodynamic stability via the Marangoni effect, the specific nonlinear dynamics and the role of interfacial deformation remain fully unclear. 
To address this, we developed a phase-field-based numerical model, validated against linear stability analysis and existing theories. 
The validated parameter space includes Schmidt number, Marangoni number, Capillary number, and the diffusivity and viscosity ratio between the two layers.
Our finding shows that solute transfer from a less diffusive layer triggers short-wave instability, governed by the critical Marangoni number, while solute transfer into a less viscous layer induces long-wave instability, controlled by the critical Capillary number.
Nonlinear simulations reveal distinct field coupling behaviors: in the diffusivity-ratio-driven instability, the spatially averaged flow intensity remains symmetric about a flat interface, while solute gradient is uneven. In contrast, in viscosity-ratio-driven instability, a deforming interface separates the two layers, with a uniform solute gradient but asymmetric spatially averaged flow intensity.
These results highlight the crucial role of diffusivity and viscosity in shaping Marangoni flows and enhance our understanding of interfacial instability dynamics.}
\end{abstract}

\end{frontmatter}


\section{Introduction}

Multicomponent fluids, comprised of two or more chemical species, are integral to a wide range of natural and industrial processes, from chemical production to material manufacturing~\citep{brennecke_ionic_2001,snell_macromolecular_2005}. 
Variations in their composition, driven by spontaneous mass transfer, mixing, or chemical reactions alter fluid properties such as surface tension, viscosity, and diffusivity.
These composition-dependent alterations drive hydrodynamic phenomena,  inducing Marangoni flows~\citep{levich1969surface,manikantan2020surfactant} and osmosis flows~\citep{velegol2016origins,shim2022diffusiophoresis}, which can substantially impact micro-flow dynamics and system behavior.

Recently, there have been several complex hydrodynamic phenomena reported in multicomponent microfluidic systems, such as spontaneous phase separation~\citep{tan2016evaporation, tan2019microdroplet, guo2021non}, self-lubrication~\citep{tan2019porous, tan2023self}, self-explosion~\citep{lyu2021explosive}, propulsion~\citep{izri2014self,jin2017chemotaxis}, Marangoni spreading and contracting~\citep{DIeter2022, chao2022liquid}, attraction and chasing~\citep{cira2015vapour,li2023oil}, and targeted-migration of microdroplets~\citep{banerjee2016soluto,tan2021two,may2022phase}, making multicomponent hydrodynamics a subject of great interest to the fluid mechanics community~\citep{maass2016swimming, lohse2020physicochemical, manikantan2020surfactant,wang2022wetting, shim2022diffusiophoresis, dwivedi2022self}. 
The phenomena in these small-scaled liquid systems, share the same fact that the coupling of multiple fields is essential to the richness of the emerging interfacial hydrodynamics. 
Thus, a comprehensive understanding of the interplay at the interface is crucial to the design of micro-flow systems.

In particular, the Marangoni effect, which predates Thomson's observation of the tears of wine effect in the 19th century, involves a complex interplay among interfacial flow fields, solute concentration fields, and the deformation of the liquid-liquid interfaces. 
This phenomenon occurs when surface-active solutes, acting as the third component, are unevenly distributed at the interface between immiscible liquids~\citep{levich1969surface}. 
These solutes influence the concentration-dependent interfacial surface tension, denoted as $\gamma$, leading to concentration gradient along the interface.
This gradient creates an imbalance that triggers non-equilibrium liquid motions, known as solutal Marangoni flow.
These emerging flows advectively transport the solutes, fostering interaction between the flow and concentration fields~\citep{manikantan2020surfactant}.
Simultaneously, the directed motion of flow towards the interface induces interfacial deformation, adding another layer of complexity to the coupling process. 
Consequently, flow field, composition field, the interfacial deformation become intricately intertwined within the solutal Marangoni flow.
An essential and fundamental inquiry pertains to the stability of this interfacial coupling.

In a groundbreaking study, Sterning and Scriven~\cite{sternling_interfacial_1959} conducted stability analysis and studied Marangoni hydrodynamic instability between two un-equilibriated fluid $without$ interfacial deformation. 
They summarized that the stability depends on the solute transfer direction, the viscosity and diffusion ratios of fluids on both sides, and the relationship between the surface tension coefficient and the solute concentration. 
Later, Reichenbach and Linde~\cite{reichenbach_linear_1981} modified the model by considering a normal deformation of the interface.
Since then, Marangoni interfacial instability has garnered significant attention, leading to various numerical and experimental investigations aimed at comparing the results obtained by \cite{sternling_interfacial_1959}, exploring criteria such as the Marangoni number for instability or oscillatory stability~\citep{Wei2006, kovalchuk_marangoni_2006,schwarzenberger_pattern_2014}, and examining instability in different flow systems~\citep{degen1998time,kalliadasis_marangoni_2003,kalogirou_capturing_2016,mokbel_influence_2017}.

The small length and short time scales in experiments pose challenges in studying the development of concentration, flow fields, and interfacial deformation, as well as their coupling and instability mechanisms~\citep{kovalchuk_marangoni_2006, shin_benard-marangoni_2016,li_marangoni_2021}.
Direct Numerical Simulation (DNS) offers a detailed exploration of these processes, but most numerical methods for studying Marangoni instability rely on sharp interface models and linear stability analysis, which struggle with large interface deformations and nonlinear effects.
Although some researchers have extended linear models to account for nonlinearity ~\citep{Wei_2005, kalogirou_nonlinear_2020}, a gap remains between these approaches and real physical systems.

In this work, we propose a phase-field-based numerical model to investigate Marangoni instability at a deformable interface, explicitly accounting for nonlinear effects.
We focus on solutal Marangoni flows driven by transverse and sustained solute transfer across an initially flat liquid-liquid interface, neglecting gravitational effects.
Section~\ref{sec:2} outlines the mathematical formulation, introducing a model that couples the convective Allen-Cahn equation, advection-diffusion equation, and Navier-Stokes equations, and linearizes the system for stability analysis.
Section~\ref{sec:3} presents the results, examining the influence of various parameters on Marangoni instability, comparing them with theoretical predictions~\cite{sternling_interfacial_1959, reichenbach_linear_1981}, and exploring nonlinear effects via DNS.
Finally, Section~\ref{sec:4} summarizes the key findings and conclusions  from our study.

\section{Problem Formulation and Methodology}
\label{sec:2}
\begin{figure}        
\centerline{\includegraphics[clip, width=13cm]{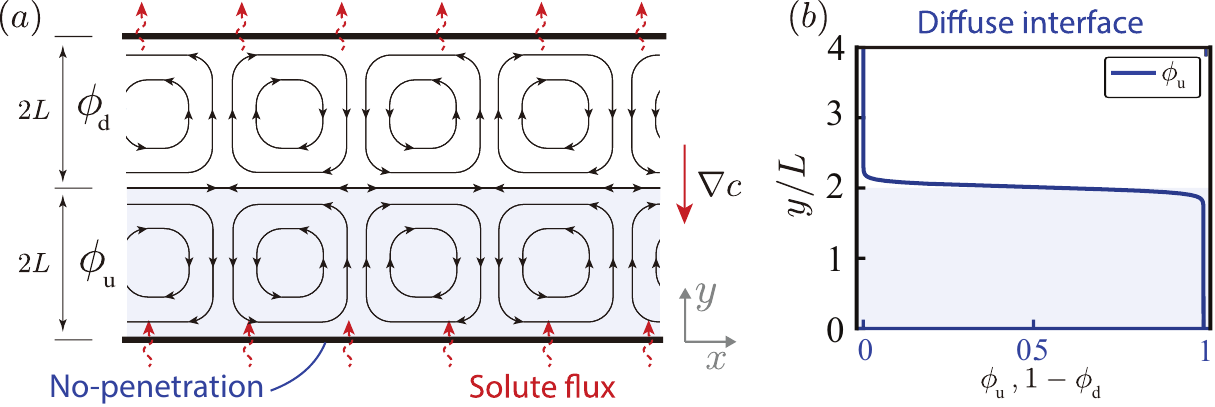}}
\caption
{Physical problem formulation with a phase-field method.
($a$) Schematic of a two-liquid layer system. 
($b$) The equilibrium state of the phase parameter distribution $\phi_{\text{u}}$ (blue line) along the $y$-direction.
}
\label{fig:ps_pc}
\end{figure}       

\subsection{Mathematical Formulation and Numerical Model}\label{sec:2.1}
We analyze a two-dimensional system with two immiscible fluids confined between infinite, flat, non-penetrating surfaces at $y=0$ and $y=4L$, as depicted in Fig.~\ref{fig:ps_pc}. 
The interface at $y=2L$ separates the bottom fluid $\phi_{\text{u}}$ from the top fluid $\phi_{\text{d}}$, where the subscripts `u' and `d' represent the upstream and downstream phases, respectively.
A downward concentration gradient $\nabla c$ exists perpendicular to the interface.
No-slip, non-penetrating walls impose concentration boundary conditions:  $c_{y=4L}=0$ and $c_{y=0}= 4G_c L$, where $G_c$ is characteristic concentration gradient, while periodic boundary conditions apply laterally.
Though transverse mass transfer does not initially induce a concentration difference along the interface, any perturbation creates solute imbalance, triggering solutal Marangoni flows and forming two-dimensional roll cells \citep{sternling_interfacial_1959}.

The primary focus of this study is the subsequent behavior of the triggered solutal Marangoni flows, specifically under various conditions. 
For unstable conditions, the triggered development of flow is amplified and sustained by the continuous mass transfer, whereas stable conditions result in the dissipation of the flows. 
To explore this phenomenon, a combination of DNS and stability analysis is employed.
\\

To track interfacial deformation, we use a diffuse interface in the numerical model.
The phase-field method is applied to simulate the evolution of the flow field, composition field, and interfacial deformation.
An order parameter $\phi$ is used to track the interface, and the conservative Allen-Cahn equation, incorporating the advection term~\cite{chiu2011conservative,aihara2019multi}, is applied to capture the deformable interface in a flow field $\boldsymbol{u}$.
For the bottom $\phi_{\text{u}}$ and top $\phi_{\text{d}}$ liquid phases, we define two phase field equations, i.e.,
\begin{equation}
\label{eq:phasefield}
\frac{\partial \phi_{i}}{\partial t^*}+\nabla^{*} \cdot\left(\phi_{i} \boldsymbol{u^{*}} \right)=\frac{1}{Pe_\phi} \left\{\nabla^{*} \cdot\left[\nabla^{*} \phi_{i}-B^* \phi_{i}\left(1-\phi_{i}\right) \frac{\nabla^{*} \phi_{i}}{\left|\nabla^{*} \phi_{i}\right|}\right]+a^*_i \right\}, 
\end{equation}	   
where $i=\text{u} \, \text{or} \, \text{d}$, and the superscript $*$ denotes dimensionless quantities. 
The Péclet number of phase field, $Pe_\phi$, characterizes the relative mass transfer of $\phi$ due to convective deformation and the anti-diffusion, which restore equilibrium when tracking the interfacial deformation. 
The equilibrium distribution of $\phi$ is given by $ \phi_{\mathrm{eq}}=\frac{1}{2}\left[1-\tanh \left(B^*(y^*-2)/2\right)\right]$, as depicted by the solid blue line in Fig.~\ref{fig:ps_pc}$b$.        
In the simulation, the constant $B^*=4/\delta^* \operatorname{artanh}(1-2\delta_{cr})$, where $\delta^*$ is the interface thickness, and ${\delta}_{cr}=0.05$ defines the numerically achievable interface region as $\delta_{cr}\leq \phi \leq (1-\delta_{cr})$. 
The constraint $\phi_{\text{u}}+\phi_{\text{d}}=1$ is satisfied using Lagrange multipliers $a_\text{u}^*$ and $a_\text{d}^*$, as described in \cite{lee2015efficient}.

The fluid properties, including viscosity $\mu^*$ and solute diffusivity $D^*$, are assumed as,
\begin{equation}
\label{eq:den_vis}
\mu^* =  \phi_{\text{u}} +\frac{1}{\zeta_\mu}\phi_{\text{d}},\quad D^*  =  \phi_{\text{u}} + \frac{1}{\zeta_D}\phi_{\text{d}},
\end{equation}             
where $\zeta_\mu=\mu_{\text{u}}/\mu_{\text{d}}$ and $\zeta_D=D_{\text{u}}/D_{\text{d}}$ are viscosity and diffusivity ratios between the phases. 
The density of both layers is assumed equal, as gravitational effects are neglected.

For interfacial surface tension, we follow the approach in \cite{kim_phase-field_2012} and express it as
\begin{equation}
\label{eq:sf}
\boldsymbol{SF^*} = \frac{1}{2} \Big[\gamma^*
\left(\kappa^*_{\text{u}} \boldsymbol{n}_{\text{u}}+\kappa^*_{\text{d}} \boldsymbol{n}_{\text{d}}\right)
+ (\boldsymbol{I}_{\text{u}}-\boldsymbol{n}_{\text{u}} \boldsymbol{n}_{\text{u}}+
\boldsymbol{I}_{\text{d}}-\boldsymbol{n}_{\text{d}} \boldsymbol{n}_{\text{d}}) \cdot \nabla^* \gamma^* \Big]
W^*,
\end{equation}
where the local normal vector $\boldsymbol{n}_i=\nabla^* \phi_i/|\nabla^* \phi_i|$, and curvature $\kappa_i^*=-\nabla^* \cdot(\nabla^* \phi_i/|\nabla^* \phi_i|)$ are linked to the phase fields.
The term $\gamma^* \kappa^* \boldsymbol{n}$ represents normal stress due to capillary pressure, while
$(\boldsymbol{I}-\boldsymbol{n} \boldsymbol{n})\cdot \nabla^* \gamma^*$ represents tangential stress due to the Marangoni effect.
The weight function $W^*$ is defined as $A^* \phi_{\text{u}} \phi_{\text{d}} |\nabla^* \phi_{\text{u}}| |\nabla^* \phi_{\text{d}} |$, with $A^*=-30/B^*$.

Considering incompressible, isothermal flow with a dilute solute, the governing flow equations are,
\begin{align}
    \rho^*\left(\frac{\partial \boldsymbol{u^{*}}}{\partial t^{*}}+\boldsymbol{u^{*}} \cdot \nabla^{*} \boldsymbol{u^{*}}\right)&=\frac{Sc}{Ma}\bigg\{-\nabla^{*} p^{*}+\nabla^* \cdot\left[\mu^{*}\left(\nabla^{*} \boldsymbol{u^{*}}+(\nabla^{*} \boldsymbol{u^{*}})^{T}\right)\right]
    +\frac{\boldsymbol{S F}^{*}}{Ca}\bigg \}, \label{eq:NS}\\
    \nabla^* \cdot \boldsymbol{u}^*&=0, \label{eq:continous}
\end{align}
where the solute concentration field follows the advection-diffusion equation
\begin{equation}
\label{eq:AD}
\frac{\partial c^*}{\partial t^*}= \frac{1}{Ma}\nabla^{*}\cdot \left[D^* (\nabla^* c^*)\right]-\nabla^*\cdot (c^*\boldsymbol{u}^*). 
\end{equation}
The dimensionless numbers include the Marangoni number of the solute concentration field $Ma=UL/D_{\text{u}}$, the Schmidt number $Sc = \mu_{\text{u}}/(\rho_\text{u} D_{\text{u}})$, and the Capillary number $Ca=U\mu_{\text{u}}/\gamma_{0}$.
Although the solute is considered dilute, small concentrations can significantly alter the (interfacial) surface tension coefficient $\gamma^*$, following a simplified linear relationship $\gamma^*= 1-Ca(c^*-c^*_0)$, as described in~\citep{khossravi_solvent_1993,Picardo_2016}.
In the DNS, a global random perturbation with a magnitude of 0.0001 relative to the maximum of solute concentration is applied throughout the system.
To characterize the solutal Marangoni flow in the low Reynold number regime, we use $U= \beta |G_c|L/\mu_{\text{u}}$ as the characteristic velocity, where $\beta \coloneqq \text{d}\gamma/\text{d}c$ is a positive coefficient, indicating the sensitivity of surface tension to concentration changes.
The definition of all dimensionless quantities are given in \ref{sec:Nondimensionlize}.

\subsection{{Linear stability analysis}}
\label{subsec:simpliy_linearize}

To investigate the onset of instability induced by perturbed solute concentration gradient at the interface, we perform a linear stability analysis of the flow.
In this framework, the state vector $\boldsymbol{q}^*=(\boldsymbol{u}^*, p^*, c^*, \phi_{\text{u}})^T$ is decomposed into a base state $\boldsymbol{Q}^b=(\boldsymbol{U}^b, P^b, C^b, \phi_{\text{u}}^b)^T$ and a small perturbation $\boldsymbol{q}^\prime=(\boldsymbol{u}^\prime, p^\prime, c^\prime, \phi_{\text{u}}^\prime)^T$, such that 
$\boldsymbol{q}^* =\boldsymbol{Q}^b + \boldsymbol{q}^\prime$. 
The base state and perturbation details are provided in equations~(\ref{eq:base_state}) and (\ref{eq:normal_mode}). 

The linearized system (\ref{eq:linear_eq}) admits normal mode solutions of the
form $\boldsymbol{q}' = \tilde{\boldsymbol{q}}(y^*) \text{e}^{\text{i}kx^*-\text{i}\omega t^*} + \text{c.c.}$, where $\text{i}$ is the imaginary unit, $k$ is the real-valued wavenumber, and
$\omega=\omega_{r} + \text{i} \omega_i$ is the complex frequency.
Here, $\omega_i$ is the linear growth rate, $\omega_r$ is the oscillation frequency, and variables with tilde $\tilde{}$ are related to Fourier space (detailed in~\ref{sec:linear_analysis}).
The linear equation system in Fourier space is presented in equation~(\ref{eq:linear_eq_Fourier}). 
By denoting the state vector of the eigenfunctions as $\tilde{\boldsymbol{q}}=(\tilde{u},\tilde{v},\tilde{p},\tilde{c},\tilde{\phi}_\text{u})^T$, the system can be compactly written as a generalized eigenvalue problem,
\begin{equation}
\omega \widetilde{\boldsymbol{M}} \tilde{\boldsymbol{q}}=  \widetilde{\boldsymbol{L}} \tilde{\boldsymbol{q}},
\end{equation}
where $\widetilde{\boldsymbol{M}}$ and $\widetilde{\boldsymbol{L}}$ are matrices derived from the linearized equations.
The explicit forms of these matrix elements can be found in~\ref{sec:linear_analysis}.
In this analysis, $\omega_i>0$ indicates that the Marangoni flow is linearly unstable to infinitesimal disturbances, while $\omega_i<0$ means the flow is linearly stable. 
$\omega_i=0$ corresponds to a neutral state, and $\omega_r \ne 0$ signifies oscillations in the system's energy during its evolution.

\section{Results}
\label{sec:3}

\subsection{Physical Parameter Space and Diffuse Interface Effect}
\label{sec:3.1}

\begin{figure}[h!]
\centerline{\includegraphics[width=\linewidth]{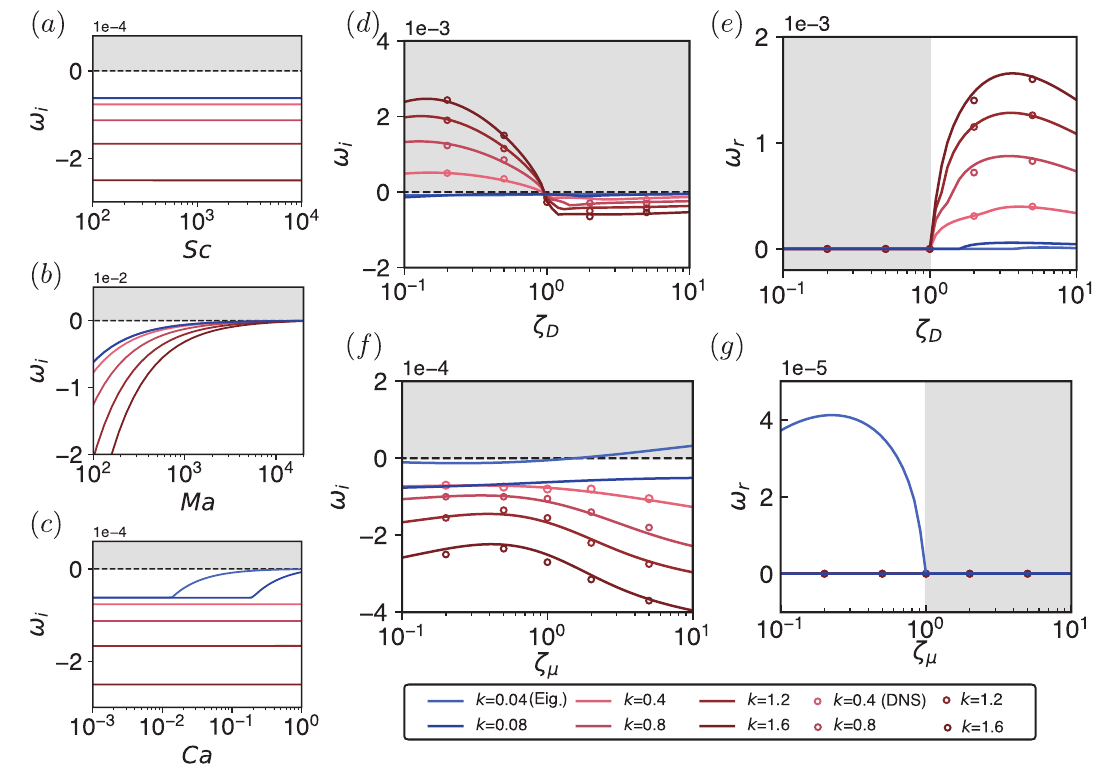}}
\caption{ Influence of physical parameters $Sc$, $Ma$, $Ca$, $\zeta_D$, and $\zeta_\mu$ on the Marangoni interfacial instability, based on linear stability analysis (Eqns. 2.17) depicted by solid lines, and DNS computations (Eqns. 2.10) represented by circular dots. 
Bluish colors correspond to smaller wave numbers $k = 0.04,0.08$, while reddish colors represent larger wave-numbers $k = 0.4,0.8,1.2,1.6$.
The investigated regime includes: ($a$) $Sc \in [10^2,10^4]$ with $Ma = 10^4$, $Ca=10^{-2}$, ($b$) $Ma \in [10^2,2\times 10^4]$ with $Sc = 10^3$, $Ca=10^{-2}$, and ($c$) $Ca \in [10^{-3},1]$ with $Sc = 10^3$, $Ma=10^4$.
For $\zeta_\mu = 1$, $\zeta_D = 1$, the growth rates $\omega_i$ for various wave numbers are negative, indicating stable regimes (white regions). 
($d$-$e$) When $Sc = 10^3$, $Ma = 10^4$, and $Ca = 10^{-2}$ with $\zeta_\mu=1$, if $\zeta_D<1$, the growth rates $\omega_i>0$, indicating an unstable regime (gray region); if $\zeta_D>1$, the growth rates $\omega_i\leq0$, but the oscillation frequency $\omega_r>0$, indicating oscillatory decay (white region). 
($f$-$g$) For the same $Sc$, $Ma$, and $Ca$ with $\zeta_D=1$, if $\zeta_\mu \in [0.1,10]$, the system remains in a stable regime ($\omega_i\leq0$); however, if $\zeta_\mu>1$, the system becomes unstable. 
}
\label{fig:parameter}
\end{figure}

To investigate the effect of physical parameters on system stability, we conducted a parameter sweep using linear stability analysis, capitalizing on its computational efficiency.
The five key parameters examined were the Schmidt number $Sc$, Marangoni number $Ma$, Capillary number $Ca$, viscosity ratio $\zeta_\mu$, and diffusivity ratio $\zeta_D$. 
The resulting parameter space for these variables is shown in Fig.~\ref{fig:parameter}. 
The diffuse interface thickness set to 0.02 (details is given below).
The results from DNS are also included, with circular markers representing DNS data and solid lines indicating eigenfunction solutions, color-coded by wavenumber $k$.
The coincidence of dots and lines demonstrates mutual validation between the DNS method and linear stability analysis during the early linear stage.

For two liquid layer with equal dynamic viscosity and diffusivity, $\zeta_\mu=1$ and $\zeta_D=1$, the Marangoni flow remains when varying $Sc$, $Ca$, or $Ma$ (Figs~\ref{fig:parameter}a-c). 
However, when $\zeta_D$ is varied from 0.1 to 10 while keeping $Sc = 10^3$, $Ma = 10^4$, and $\zeta_\mu=1$, we observe that $\omega_i>0$ within the range $\zeta_D \in (0.1, 1)$ (Figs~\ref{fig:parameter}d-e). This indicates continuous growth of the Marangoni flows due to solute transfer into a higher diffusivity layer, signifying an unstable regime. 
In contrast, for $\zeta_D \in (1, 10)$, $\omega_i<0$ and $\omega_r \ne 0$ suggest an oscillatory decay of the flow, marking a table regime dominated by damped oscillations.
When $\zeta_\mu$ is varied, instead of $\zeta_D$, from 0.1 to 10, $\omega_i>0$ in the range $\zeta_{\mu} \in (1, 10)$ (Figs~\ref{fig:parameter}f-g), indicating that solute transfer out of a more viscous layer induces instability.
The effects of $\zeta_\mu$ and $\zeta_D$ on system stability, by employing a phase-field-based model, agree with that the pioneer work by~\cite{sternling_interfacial_1959} with a non-deformable interface, and the work by~\cite{reichenbach_linear_1981} with a normal deformation of the interface.

\begin{figure}[h]
    \centerline{\includegraphics[width=12cm]{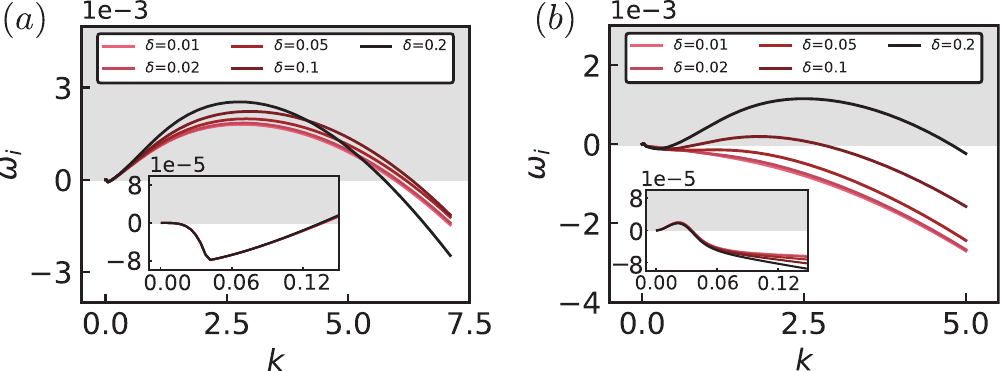}}
    \caption{ Instability at different wave numbers and diffuse interface effect.
($a$) In the diffusivity-ratio-driven instability case with $Sc = 10^3$, $Ma = 10^4$, $Ca = 10^{-2}$, $\zeta_\mu = 1$, and $\zeta_D = 0.5$, a persistent short-wave instability is observed as the dimensionless interfacial thickness $\delta^*$ varies from 0.01 to 0.2. This indicates that interfacial thickness has no significant effect on the diffusivity-related short-wave instability.
($b$) In the viscosity-ratio-driven instability case with $Sc = 10^3$, $Ma = 10^4$, $Ca = 10^{-2}$, $\zeta_\mu = 5$, and $\zeta_D = 1$, a persistent long-wave instability is observed as $\delta^*$ varies from 0.01 to 0.2. However, a short-wave instability emerges when the interfacial thickness becomes larger.
    }
    \label{fig:thickness}
\end{figure}

The diffuse interface is an inherent feature of the phase-field method.
Since the interface thickness $\delta^*$ is a crucial physical parameter in simulations~\citep{demont_numerical_2023}, it is essential to determine its value before analyzing the Marangoni instability. 
We employed linear stability analysis to evaluate the influence of interface thickness $\delta^*$ on the growth rate $\omega_i$ across different wavenumbers $k$.
This was done for both a diffusivity-ratio-driven instability case ($\zeta_\mu = 1$, $\zeta_D = 0.5$) in Fig.~\ref{fig:thickness}$a$, and a viscosity-ratio-driven instability case ($\zeta_\mu = 5$, $\zeta_D = 1$) in Fig.~\ref{fig:thickness}$b$, with $Sc = 10^3$, $Ma = 10^4$, and $Ca = 10^{-2}$.
In both cases, the growth rate curves converge as $\delta^*$ approaches zero, indicating a transition toward the sharp interface limit.
Differences between the curves are negligible when $\delta^* \le 0.05$, and we therefore selected $\delta^* = 0.02$ for the parameter sweep and subsequent calculations to reduce computational cost.
However, it is important to note that the interface thickness influences the onset of Marangoni instability.

In the diffusivity-ratio-driven instability (Fig.~\ref{fig:thickness}$a$), a short-wave instability (wave numbers $k>0.14$) consistently occurs across the range of interface thickness from 0.01 to 0.2.
The range of unstable wavenumbers depends on the thickness.
In the viscosity-ratio-driven instability case (Fig.~\ref{fig:thickness}$b$), we observed a long-wave instability (wave numbers $k<0.03$) that is independent of the interface thickness, while a short-wave instability emerges as the thickness increases ($\delta^* \ge 0.1$).
The underlying mechanism of the diffuse interface effect remains unclear.
Although the sharp interface assumption is common in multiphase flow, physically, there is a continuous variation of density across the interfacial region, with a thickness of $\sim$nm between the two bulk phases in equilibrium~\cite{yang1996method}.
This suggests that the diffuse interface effect cannot always  be neglected at submicroscopic scales, and it may play a dominant role in miscible solution systems~\cite{vorobev2014dissolution}.

\subsection{Neutral Stability and the Critical Ma and Ca Numbers}
\label{sec:3.2}
\begin{figure}[h]
    \centerline{\includegraphics[width=12cm]{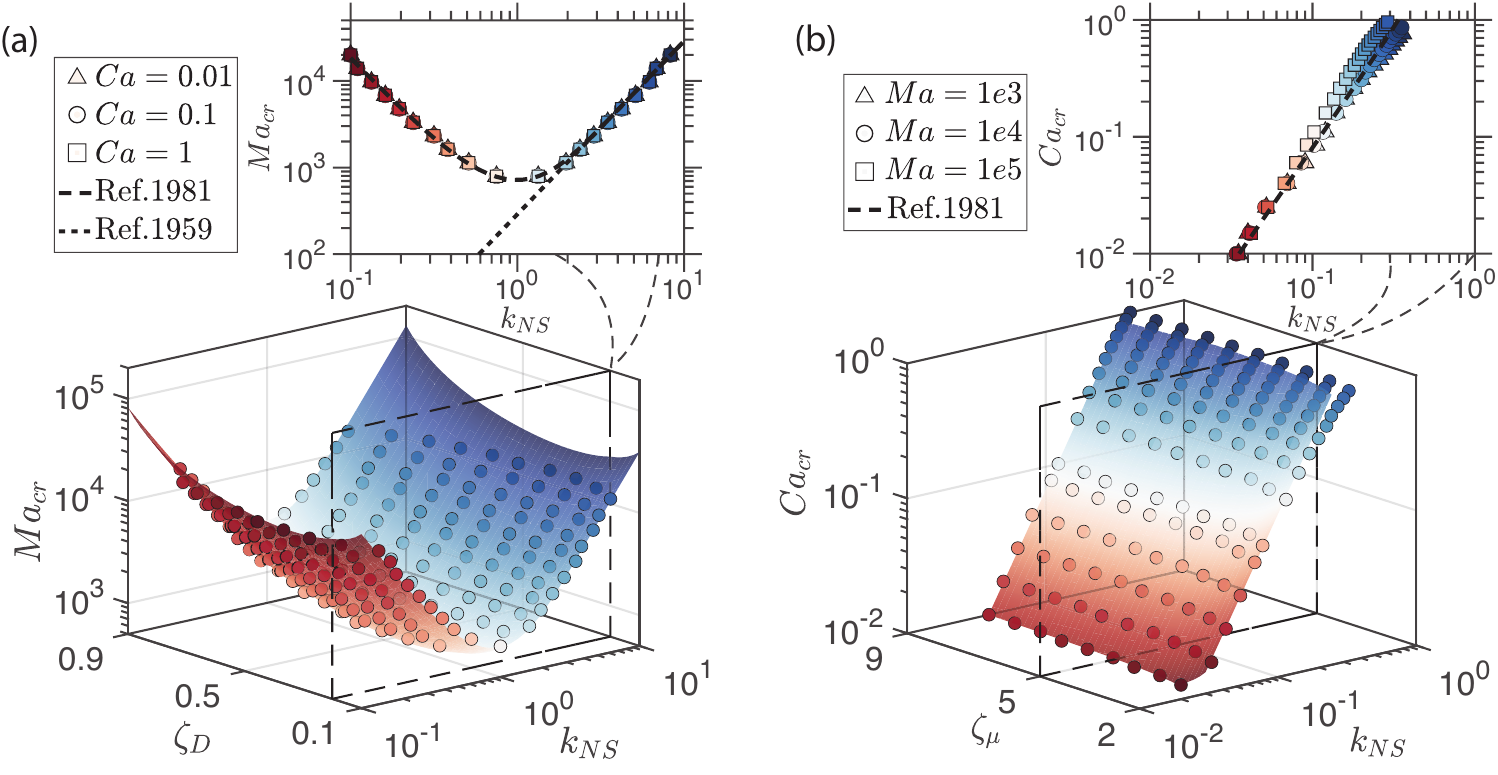}}
    \caption{ Neutral stability diagrams.
    ($a$) $Ma_{cr}$ for varying diffusivity ratios $\zeta_D \in [0.1, 0.9]$ with $\zeta_\mu = 1$, and ($b$) $Ca_{cr}$ for varying viscosity ratios $\zeta_\mu\in [2, 9]$ with $\zeta_D = 1$, both at $Sc = 10^3$ and $\delta = 0.02$. 
Dots represent calculated values, while the aligned surfaces show theoretical predictions from the literature~\cite{reichenbach_linear_1981}. 
Inset ($a$) indicates that neutral stability curves at $\zeta_D = 0.2$ for $Ca = 10^{-2}$, $10^{-1}$, and $1$ collapse onto the theoretical predication, suggesting that the $Ma_{cr}$ governs diffusivity-related instability.  
Inset ($b$) shows that the neutral stability curves at $\zeta_\mu = 5$ for $Ma = 10^3$, $10^4$, and $10^5$ collapse onto the theoretical predication  for small $Ca_{cr}$, indicating that $Ca_{cr}$ controls viscosity-related instability when surface tension effect dominates.
}
    \label{fig:Macr}
\end{figure}

The neutral stability boundary, marking the transition from stability ($\omega_i<0$) to instability ($\omega_i>0$), is defined by the neutrally stable wavenumber $k_{NS}$ and determines the onset of the Marangoni-driven flows~\cite{borcia_phasefield_2003,kovalchuk_marangoni_2006,lopez_de_la_cruz_marangoni_2021}. 
To capture the critical conditions of flow instability, we focus on the Marangoni number $Ma_{cr}=\beta |G_c| L^2/(\mu_{\text{u}}D_\text{u})$ and Capillary number $Ca_{cr}= \beta |G_c| L/\gamma_0$, which describes the balance between surface tension gradients, viscosity, and inertial forces.

Sternling \& Scriven~\cite{sternling_interfacial_1959} provided a classical prediction of $Ma_{cr}$ for non-deformable interfaces. 
They showed that instability occurs when $Ma>Ma_{cr}$, and expressed the relation between $Ma_{cr}$ and $k_{NS}$ for a non-deformable interface as 
\begin{equation}
    \label{eq:crMa}
    Ma_{cr} =8 a_{\zeta_D}\frac{\left(\zeta_D^{-1}+1\right)\left(\zeta_\mu^{-1}+1\right)}{\left(\zeta_D-1\right)}  {k_{NS}^2},
\end{equation}
where $\alpha_{\zeta_D}$ serves as a correction factor, since the characteristic concentration gradient $|G_c|$ is defined here as $(c_{y=4L}-c_{y=0})/(4L)$, rather than $(c_{y=2L}-c_{y=0})/(4L)$ (upstream layer) as used in the reference~\cite{sternling_interfacial_1959}.
However, Reichenbach \& Linde~\cite{reichenbach_linear_1981} accounted for interfacial  deformation and proposed a modified model. 
Applying their model to our system, we obtained the expression of $Ma_{cr}$ and $Ca_{cr}$, and the simplified formulations of $Ma_{cr}$ for $\zeta_\mu = 1$ and an introduced formulation of $Ca_{cr}$ for $\zeta_D=1$ are,
\begin{equation}
    \label{eq:crMa2}
    M a_{cr}=2a_{\zeta_D}\frac{\zeta_D^{-1}+1}{\zeta_D-1} \frac{ H_{\text{u}}\left({1+E_{\text{d}}}\right)}{{L_\text{d}}},
\end{equation}
\begin{equation}
    \label{eq:crMa2_2}
    Ca_{cr}=2{k_{NS}^2} \frac{H_\text{u}+\zeta_\mu^{-1} H_\text{d}}
    {P_\text{u}-\zeta_\mu^{-1}P_\text{d}},
\end{equation}
where $H$, $E$, $L$, and $P$ are function of $k_{NS}$, and the detailed symbol definitions and derivations of these expressions are provided in~\ref{appA}.

Fig.~\ref{fig:Macr}$a$ shows the critical Marangoni number $Ma_{cr}$ as a function of the diffusivity ratio ($\zeta_D\in[0.1, 0.9]$) for a fixed viscosity ratio ($\zeta_\mu=1$). 
Dots represent calculated values, while the 3D surface and 2D dashed line correspond to Reichenbach and Linde's predication~\cite{reichenbach_linear_1981} (Eq.~\ref{eq:crMa2}).
The inset of Fig.~\ref{fig:Macr}$a$ highlights that for $\zeta_D=0.2$, the neutral stability curves at various Capillary numbers ($Ca = 0.01, 0.1, 1$) converge onto the dashed line, indicating that $Ma_{cr}$ governs diffusivity-ratio-driven instability.
Sternling \& Scriven's non-deformable interface model~\cite{sternling_interfacial_1959} (Eq.~\ref{eq:crMa}), represented with dotted line, is valid only for $k_{NS}>2$ and fails for longer wavelengths, underscoring the importance of considering interfacial deformation.
We note that $Ma_{cr}$ exhibit
power-law relationships with $k_{NS}$, including an exponent of -2 when $k_{NS}\ll1$ and 2 when $k_{NS}\gg1$.

Similarly, Fig.~\ref{fig:Macr}$b$ presents $Ca_{cr}$ as a function of viscosity ratio ($\zeta_\mu \in [1, 9]$) with a fixed diffusivity ratio ($\zeta_D=1$).
The 3D surface and 2D dashed line from Reichenbach \& Linde's model~\cite{reichenbach_linear_1981}  (Eq.~\ref{eq:crMa2_2}) align with the calculated dots.
The inset shows that for $\zeta_\mu=5$, the neutral stability curves increases monotonically at various $Ma$ values ($10^3, 10^4, 10^5$), with deviations emerging as $Ca$ increases, indicating a reduced influence of surface tension.
We note that $Ca_{cr}$ and $k_{NS}$ can be fitted with a power-law relationship of $Ca_{cr} \sim k_{NS}^2$.


\subsection{Interactions Among Multiple Fields and Nonlinear Effects}
The DNS results offer insights into the coupling processes among the flow field, phase field, and solute field during the linear development stage, as well as the saturation behavior in the later stages where nonlinear effects dominate.
Specifically, we examine a diffusivity-ratio-driven unstable case ($\zeta_D=0.2, \zeta_\mu=1$) and a viscosity-ratio-driven unstable case ($\zeta_D=1, \zeta_\mu=10$), with parameters set to $Sc=10^3$, $Ma=10^4$, $Ca=10^{-1}$, and $Re=Ma/Sc=10$. 

\begin{figure}[h!]
    \centerline{\includegraphics[width=12cm]{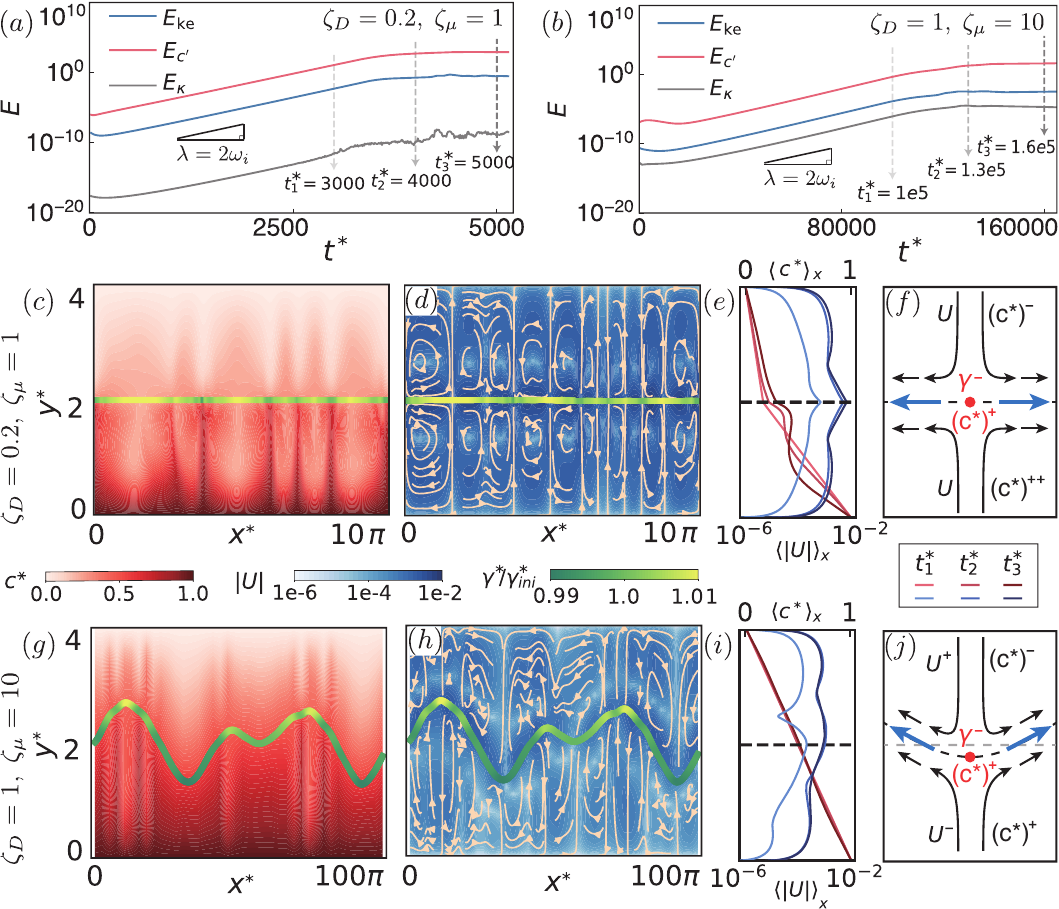}}
    \caption{ Nonlinear effects on the instability evolution.
    $(a)$ Energy evolution $E=(E_\text{ke}, E_{c^\prime}, E_{\kappa})$ for diffusivity-ratio-driven instability ($\zeta_D=0.2,\,\zeta_\mu=1$) and $(b)$ for viscosity-ratio-driven instability ($\zeta_D=1,\,\zeta_\mu=10$), based on DNS results with $Sc=10^3$, $Ma=10^4$, and $Ca=10^{-1}$.
The energy grows linearly at rate $\lambda$, consistent with  linear stability predications ($2\omega_i$), before saturating in the nonlinear regime.
For the diffusivity-case, $(c)$ and $(d)$ show snapshots of the concentration field $c^*$ (red) and flow intensity field $|U|$ (blue) at $t_3^*=5000$, with the greenish lines ($\gamma^*/\gamma^{b}$ color-coded) marking the deformable interface and orange arrows showing streamlines. 
$(e)$ show x-averaged profile $\langle c^*\rangle_x$ and $\langle|U| \rangle_x$ at $t_1^* = 3000$, $t_2^* = 4000$, and $t_3^* = 5000$. 
$(f)$ explain the coupling effect among concentration field, flow field, and phase field. 
$(g-j)$ depict the corresponding fields for the viscosity-case at $t_1^* = 10^5$, $t_2^* = 1.3\times  10^5$, and $t_3^* = 1.6\times10^5$.  (Movies S1-S4)
}
    \label{fig:DNS}
\end{figure}

To illustrate the instability development across multiple fields, we defined a set of energy measures $E$, which includes solute concentration $E_{c^\prime}$, kinetic energy $E_\text{ke}$, and interfacial deformation $E_{\kappa}$, as,
\begin{equation}
\label{eq:ek}
E^{}_{c^\prime}=\int_{0}^{l_x}\int_{0}^{l_y}c^{\prime 2} \mathrm{d}y^* \mathrm{d}x^*,\quad
E^{}_\text{ke}=\int_{0}^{l_x}\int_{0}^{l_y}\boldsymbol{u}^{\prime} \cdot \boldsymbol{u}^{\prime} \mathrm{d}y^* \mathrm{d}x^*,  \quad  
E^{}_{\kappa}={\int_{0}^{l_x} \bar{\kappa}^2 \mathrm{d}x^*},
\end{equation}
where the integration domain is the computational area $[0, \ell_x]\times [0, \ell_y]$.
Here, $\boldsymbol{u}^{\prime}=\boldsymbol{u}^*$ since the base state velocity field is $\boldsymbol{U}^b=\boldsymbol{0}$ in the simulation. 
The calculation of $\bar{\kappa}$ follows the method used in references~\citep{brackbill_continuum_1992,popinet_numerical_2018}.

Figure~\ref{fig:DNS}$a$ and $b$ present the evolution of the energy set $E(E_{c^\prime}, E_\text{ke}, E_{\kappa})$ for the diffusivity-ratio-driven instability and the viscosity-ratio-driven instability, respectively.
Initially, the three energy components exhibit linear growth at an identical rate, $\lambda$, aligning with linear stability predications of $2\omega_i$, which confirms that the early stages of instability are dominated by linear effects.
The growth rate for $E_{\kappa}$, denoted as $\lambda_\kappa=2\omega_i$, is detailed in~\ref{appB}. 
As the instability develops, the growth of energy gradually decelerates, eventually reaching saturation.
This transition marks the increasing influence of nonlinear dynamics and the interaction between multiple fields.

In the diffusivity-ratio-driven case, the interface remains flat during nonlinear development, as indicated by the greenish solid line in the concentration field $c^*$ (Fig.~\ref{fig:DNS}$c$, MovieS1) and the flow intensity field $|U|=\sqrt{\boldsymbol{u}^*\cdot \boldsymbol{u}^*}$ (Fig.~\ref{fig:DNS}$d$, MovieS2) at $t^*_3=5000$.
The greenish color shows the variation in interfacial surface tension $\gamma^*$, normalized by its initial value $\gamma^b$ from the base concentration state.
The higher diffusivity in the upper layer ($\phi_\text{d}$) results in a more uniform solute distribution, as demonstrated by the initial x-averaged concentration profile $\langle c^* \rangle_x$ (reddish lines, Fig.~\ref{fig:DNS}$e$).
Meanwhile, the x-averaged flow intensity profile $\langle |U| \rangle_x$ (bluish lines, Fig.~\ref{fig:DNS}$e$) maintains symmetric about the flat interface throughout the development.  
As illustrated in Fig.~\ref{fig:DNS}$f$, lower concentration solute $(c^*)^-$ from the upper layer and much higher concentration solute $(c^*)^{++}$ from the bottom layer are convected toward the interface with equal intensity $U$, increasing the local concentration $(c^*)^+$ at the flat interface and reducing surface tension $(\gamma)^-$.
This concentration-flow coupling drives Marangoni flow (blue arrows), enhancing convective rolls.
Thus, the Capillary number is not a critical parameter for instability in this scenario with a flat interface. Instead, the instability is primarily governed by the Marangoni number, which represents the ratio of Marangoni-driven mass transfer to diffusion (Figure~\ref{fig:Macr}a).
The linear solute distribution in the bottom layer with less diffusivity is gradually disrupted by the enhanced convection.
However, as the convective flow intensifies, it eventually saturates due to viscous dissipation.

In contrast, the viscosity-ratio-driven case, with equal diffusivity layers, results in interface deformation during nonlinear development, as depicted in Figures~\ref{fig:DNS}$g$ (Movie S3) and \ref{fig:DNS}$h$ (Movie S4) at $t^*_3=1.6\times10^5$. 
The equal diffusivity in both layers initially produces a uniform slope in the x-averaged solute profile, but the x-averaged flow intensity profile becomes asymmetric due to the viscosity mismatch, leading to interface deformation (Fig.~\ref{fig:DNS}$i$).
As illustrated in Fig.~\ref{fig:DNS}$j$, the deformation amplifies the flow intensity, pushing the interface toward regions with higher solute concentration $(c^*)^+$ and away from regions with lower solute concentration $(c^*)^-$.
This concentration-flow-deformation coupling drives Marangoni flow (blue arrows), enhancing convective rolls. 
Since the interface is deforming, the Capillary number becomes critical to the instability, particularly at small Capillary numbers, where the effect of the Marangoni number is negligible (Figure~\ref{fig:Macr}b).
As with diffusivity-ratio case, the convective flow eventually saturates due to viscous dissipation.

\section{Conclusions}
\label{sec:4}
This study investigated Marangoni interfacial instability in a two-liquid-layer system with constant solute transfer using analytical and numerical methods.
We developed and validated a phase-field-based numerical model against linear stability analysis and prior studies~\cite{sternling_interfacial_1959, reichenbach_linear_1981}.
This approach advances the understanding of Marangoni instability, extending it to nonlinear regimes involving deformable interfaces.

Our parameter sweep confirmed that the diffusivity ratio $\zeta_D$ and viscosity ratio $\zeta_\mu$ are critical for stability, with $\zeta_D<1$ and $\zeta_\mu>1$ leading to instability.
For equal diffusivity and viscosity ratios, the Schmidt number ($Sc$), Marangoni number ($Ma$), and Capillary number ($Ca$) do not influence stability within the examined parameter space.
Neutral stability analysis identified the critical Marangoni ($Ma_{cr}$) and Capillary ($Ca_{cr}$) numbers, marking the onset of instability. 
The explored parameter space reveals that diffusion-driven instability correspond to the short-wave modes (wave numbers $k>0.14$), while viscosity-driven instability correspond to long-wave modes (wave numbers $k < 0.03$).

In nonlinear simulations, we examined cases with $\zeta_D = 0.2$ and $\zeta_\mu = 10$. 
Both exhibited initial linear growth rates consistent with linear stability predictions, ultimately reaching saturation. 
In the nonlinear regime, we propose two potential mechanisms driving the instability.
For diffusivity-ratio-driven case ($\zeta_D=0.2$), equal flow intensity across the interface maintains a relatively flat interface, but the concentration gradient difference between the layers promotes Marangoni flow, amplifying the instability. 
In contrast, for viscosity-ratio-driven case ($\zeta_\mu = 10$), the flow intensity gradient causes the interface to bend toward regions of higher and lower concentration, further intensifying the Marangoni flow and enhancing the instability.

Our findings highlight the differences between the two types of instability, demonstrating how viscosity and diffusivity ratios shape Marangoni dynamics, with interfacial deformation being key to nonlinear behavior. 
These insights extend previous work and underscore the nonlinear effect, emphasizing the critical interplay between surface tension, viscosity, and solute transport in Marangoni-driven flows.

\section*{Acknowledgement}
H.T. acknowledges this work is supported by the National Natural Science Foundation of China (No. 12102171), Shenzhen Fundamental Research Program (No. 20220814180959001), and Guangdong Basic and Applied Basic Research Foundation (No. 2024A1515010614).
M.Z. acknowledges the financial support of a Tier 1 grant from the Ministry of Education, Singapore (WBS No. A-8001172-00-00). 
D.W. is supported by a PhD scholarship (No. 201906220200) from the China Scholarship Council and an NUS research scholarship.

\newpage
\appendix
\section{Non-dimensionlization of governing equations}
\label{sec:Nondimensionlize}
We adopt the following scalings to non-dimensionalize the governing equation system,                 
\begin{equation}
\label{eq:dimensionless}
\begin{split}
&\boldsymbol{x}^{*}=\frac {\boldsymbol{x}}{L}, \quad \boldsymbol{u}^{*}=\frac {\boldsymbol{u}} {U}, \quad t^{*} = \frac {t} {L/U},\quad  \nabla^{*}=L\nabla,  \quad  p^{*}=\frac{p }{P},    \quad  \delta^{*}=\frac{\delta}{L}, \\
&c^*=\frac{c}{G_cL}, \quad \gamma^* = \frac{\gamma}{\gamma_0}, \quad A^* = \frac{A}{L},  \quad B^* = {L}{B}, \quad \kappa^*=L\kappa, \\
&\mu^{*}=\frac {\mu} {\mu_{\text{u}}}=\phi_{\text{u}}+\frac{1}{\zeta_\mu} \phi_{\text{d}}, \quad D^*=\frac{D}{D_{\text{u}}}=\phi_{\text{u}}+\frac{1}{\zeta_D} \phi_{\text{d}}, \quad \rho^* = \frac{\rho}{\rho_{\text{u}}}.
\end{split}
\end{equation}  
Here, $L$, $|G_cL|$, $\gamma_0$, $\mu_{\text{u}}$, $D_{\text{u}}$, and $\rho_{\text{u}}$ are the characteristic scales of length, concentration, interfacial surface tension, dynamic viscosity, solute diffusivity, and density, respectively. 
$\zeta_\mu=\mu_{\text{u}}/\mu_{\text{d}}$ and $\zeta_D=D_{\text{u}}/D_{\text{d}}$ are the resulting ratios of dynamic viscosity and diffusivity. 
To characterize the solutal Marangoni flow in the low Reynold number regime, we use $U= \beta |G_c L|/\mu_{\text{u}}$ as the characteristic velocity, and define the characteristic pressure by $P = \mu_{\text{u}}U/L$. 
The other dimensionless number are shown as, $Sc = \mu_{\text{u}}/(\rho_{\text{u}} D_{\text{u}})$, $Ma=UL/D_{\text{u}}$, $Ca=U\mu_{\text{u}}/\gamma_{0}$, $Pe_\phi=UL/M_{\phi}$, and $M_\phi$ is the interfacial mobility. 
According to~\cite{jacqmin_calculation_1999} and~\cite{demont_numerical_2023}, the $M_\phi$ is related to the $\delta^*$, which is recommended that $M_\phi \propto \delta^{* \alpha}$ and $\alpha \in (0,3)$, here we choose $\alpha=1$. 
\section{Linear stability analysis}
\label{sec:linear_analysis}
We decompose the state vector $\boldsymbol{q}^*=(\boldsymbol{u}^*, p^*, c^*, \phi_{\text{u}})^T$ into a base state $\boldsymbol{Q}^b=(\boldsymbol{U}^b, P^b, C^b, \phi_{\text{u}}^b)^T$ plus a small perturbation $\boldsymbol{q}^\prime=(\boldsymbol{u}^\prime, p^\prime, c^\prime, \phi_{\text{u}}^\prime)^T$ as
\begin{equation}
\label{eq:linearisation}
\boldsymbol{u}^* =\boldsymbol{U}^b + \boldsymbol{u}', \ \ \ \ p^* = P^b + p',  \ \ \ c^* = C^b + c',  \ \ \ \phi_{\text{u}} = \phi_{\text{u}}^b + \phi'_{\text{u}}.
\end{equation}
In terms of the phase field variables, due to the constraint $\phi_{\text{u}} + \phi_{\text{d}} =1$, only one phase field equation needs to be solved. 
In the present implementation, we choose to solve the equation of $\phi_{\text{u}}$.

The base state is the solution to the equation system at steady state and it can be derived analytically
\begin{align}
\label{eq:base_state}
\boldsymbol{U}^b=\boldsymbol{0}, \ \ \ \ &P^b=\text{const.},  \ \ \ C^b= \frac{c^*(4)-c^*(0)}{\int_0^{4} 1/D^{b}\,dy}\int_0^{y^*} \frac{1}{D^b}\,dy^* + c^*(0), \notag \\ &\phi_{\text{u}}^b=\phi_{\mathrm{eq}}=\frac{1}{2}\left[1-\tanh \left(B^*(y^*-2)/2\right)\right].
\end{align}
Linearization of the nonlinear equation system is done by substituting the decomposition eq.~\eqref{eq:linearisation} and then subtracting the corresponding steady base-state equations and finally discarding the nonlinear terms. 
This process leads to the following linear equation system
\begin{subequations}
\label{eq:linear_eq}
\begin{align}
\frac{\partial \phi'_{\text{u}}}{\partial t^*}= -\frac{d \phi_{\text{u}}^b}{d y^*}v' + \frac{1}{Pe_\phi} \nabla^{*2} \phi'_{\text{u}} - \frac{1}{Pe_\phi} \frac{\partial^2 \phi'_{\text{u}}}{\partial x^{*2}} -\frac{1}{Pe_\phi} B^* \left[(2\phi_{\text{u}}^b -1)\frac{\partial \phi'_{\text{u}}}{\partial y^*} + 2\frac{d \phi_{\text{u}}^b}{d y^*} \phi'_{\text{u}}\right],
\end{align}	
\begin{align}
\frac{\partial c'}{\partial t^*}= &-\frac{d C^b}{d y^*}v' + \frac{1}{Ma} \left(\frac{d D^b}{d y^*}\frac{\partial c'}{\partial y^*} +D^b\nabla^{*^2}c'\right) \notag \\ &+ \frac{1}{Ma} \left[ \left(1-\frac{1}{\zeta_D}\right)\frac{d C^b}{d y^*}\frac{\partial \phi'_{\text{u}}}{\partial y^*} + \left(1-\frac{1}{\zeta_D}\right)\frac{d^2 C^b}{d y^{*2}} \phi'_{\text{u}}\right],
\end{align}      
\begin{align}
\rho^b\frac{\partial {u}'}{\partial t^{*}} = &\frac{Sc}{Ma} \left(-\frac{\partial p'}{\partial x^*} + \mu^b \nabla^{*2} {u}' + \frac{d \mu^b}{d y^*}\frac{\partial {u}'}{\partial y^*} + \frac{d \mu^b}{d y^*} \frac{\partial {v}'}{\partial x^*} \right) \notag\\
&+ \frac{Sc}{Ma}\left[\frac{1}{Ca} A^*\phi_{\text{u}}^b (1-\phi_{\text{u}}^b)\left|\frac{d \phi_{\text{u}}^b}{d y^*}\right| \frac{d \Gamma^b}{d y^*} \frac{\partial \phi'_{\text{u}}}{\partial x^*}
-\frac{\partial c'}{\partial x^*} W^b  \right],
\end{align}
\begin{align}
\rho^b\frac{\partial {v}'}{\partial t^{*}} = &\frac{Sc}{Ma} \left(-\frac{\partial p'}{\partial y^*} + \mu^b \nabla^{*2} {v}' + 2 \frac{d \mu^b}{d y^*}\frac{\partial {v}'}{\partial y^*} \right)  \notag \\ &+ \frac{Sc}{Ma}\left[\frac{1}{Ca} A^*\phi_{\text{u}}^b (1-\phi_{\text{u}}^b)\left|\frac{d \phi_{\text{u}}^b}{d y^*}\right| { \Gamma^b} \frac{\partial^2 \phi'_{\text{u}}}{\partial x^{*2}}  \right],
\end{align}	
\begin{equation}
0=\frac{\partial u'}{\partial x^*} + \frac{\partial v'}{\partial y^*}.
\end{equation}
\end{subequations}
Here, $\Gamma^b=1-Ca(C^b-c_0^*)$, $c_0^*=0$, and $W^b=A^*\phi_{\text{u}}^b (1-\phi_{\text{u}}^b)|\nabla^*\phi_{\text{u}}^b|^2$. 
For the boundary conditions, as the base state solutions $\boldsymbol{U}^b$, $C^b$ and $\phi_{\text{u}}^b$ in eq.~\eqref{eq:base_state} have already satisfied their Dirichlet boundary conditions, the perturbative variables $\boldsymbol{u}'=\boldsymbol{0}$, $c'=0$ and $\phi'_{\text{u}}=0$ should be enforced at the walls for solving eq.~\eqref{eq:linear_eq}.

The linear equation system eq.~\eqref{eq:linear_eq} admits normal mode solutions ($k \neq 0$) in the form of
\begin{align}\label{eq:normal_mode}
&\boldsymbol{u}' = \tilde{\boldsymbol{u}}(y^*) \text{e}^{\text{i}kx^*-\text{i}\omega t^*} + \text{c.c.}, \ \ \ p' = \tilde{p}(y^*) \text{e}^{\text{i}kx^*-\text{i}\omega t^*}+\text{c.c.}, \notag \\ &c' = \tilde{c}(y^*) \text{e}^{\text{i}kx^*-\text{i}\omega t^*}+\text{c.c.}, \ \ \ \phi'_{\text{u}} = \tilde{\phi}_{\text{u}}(y^*) \text{e}^{\text{i}kx^*-\text{i}\omega t^*}+\text{c.c.}.
\end{align}
Inserting eq.~\eqref{eq:normal_mode} into eq.~\eqref{eq:linear_eq} and equating the terms of the same mode (terms with $\text{e}^{\text{i}kx^*-\text{i}\omega t^*}$) results in the linear equation system in Fourier space
\begin{subequations}\label{eq:linear_eq_Fourier}
\begin{align}
-\text{i}\omega{ \tilde{\phi}_\text{u}}= &-\frac{d \phi_{\text{u}}^b}{d y^*}\tilde{v} + \frac{1}{Pe_\phi} (\frac{d^2}{d y^{*2}} - k^2) \tilde{\phi}_{\text{u}} + \\ & \frac{1}{Pe_\phi} k^2 { \tilde{\phi}_\text{u}} -\frac{1}{Pe_\phi} B^* \left[(2\phi_{\text{u}}^b -1)\frac{\partial \tilde{\phi}_\text{u}}{\partial y^*} + 2\frac{d \phi_{\text{u}}^b}{d y^*} \tilde{\phi}_\text{u}\right].
\end{align}	
\begin{align}
-\text{i}\omega{ \tilde{c}}= &-\frac{d C^b}{d y^*}\tilde{v} + \frac{1}{Ma} \left[\frac{d D^b}{d y^*}\frac{\partial \tilde{c}}{\partial y^*} +D^b\left(\frac{d^2}{d y^{*2}} - k^2\right)\tilde{c}\right] \notag \\ &+ \frac{1}{Ma} \left[\left(1-\frac{1}{\zeta_D}\right)\frac{d C^b}{d y^*}\frac{\partial \tilde{\phi}_\text{u}}{\partial y^*} + \left(1-\frac{1}{\zeta_D}\right)\frac{d^2 C^b}{d y^{*2}}\tilde{\phi}_\text{u}\right],
\end{align}
\begin{align}
-\text{i}\omega\rho^b\tilde{u} = &\frac{Sc}{Ma} \left[-\text{i}k\tilde{p} + \mu^b \left(\frac{d^2}{d y^{*2}} - k^2\right)\tilde{u} + \frac{d \mu^b}{d y^*}\frac{\partial \tilde{u}}{\partial y^*} + \text{i} k \frac{d \mu^b}{d y} { \tilde{v}} \right] \notag\\
&+ \frac{Sc}{Ma}\left[\frac{1}{Ca} A^*\phi_{\text{u}}^b (1-\phi_{\text{u}}^b)\left|\frac{d \phi_{\text{u}}^b}{d y^*}\right| \frac{d \Gamma^b}{d y^*} \text{i} k {\tilde{\phi}_\text{u}}
-\text{i}k{\tilde{c}} W^b  \right],
\end{align}
\begin{align}
-\text{i}\omega\rho^b{\tilde{v}} = &\frac{Sc}{Ma} \left[-\frac{\partial \tilde{p}}{\partial y^*} + \mu^b \left(\frac{d^2}{d y^{*2}} - k^2\right) \tilde{v} + 2 \frac{d \mu^b}{d y^*}\frac{\partial \tilde{v}}{\partial y^*} \right]  \notag \\ &+ \frac{Sc}{Ma}\left[\frac{1}{Ca} A^*\phi_{\text{u}}^b (1-\phi_{\text{u}}^b)\left|\frac{d \phi_{\text{u}}^b}{d y^*}\right| { \Gamma^b} (-k^2){\tilde{\phi}_\text{u}}  \right],
\end{align}	
\begin{equation}
0=\text{i}k{ \tilde{u}} + \frac{\partial \tilde{v}}{\partial y^*}.
\end{equation}
\end{subequations}

\section{{Numerical Method}}
\label{sec:simpliy_linearize}
We numerically solve the non-dimensionalized governing equations, using a self-developed finite-difference-method solver.
For the time derivative terms in the phase field equations and solute equations, a third order Runge-Kutta method is applied.
A fifth-order WENO method is used to solve the advection terms in the phase field equation~(\ref{eq:phasefield}) and the NS eq.~(\ref{eq:NS}), following with~\cite{aihara2019multi}.
A simple second-order central difference method is applied on the advection term in the solute eq.~(\ref{eq:AD}) because the perturbation of solute is random in space and magnitude.
For all diffusion terms, we solve them by using the second-order central difference method.
The initial condition is given by eq.~(\ref{eq:base_state}).

For the stability analysis at different modes, we perform a Fast Fourier Transform (FFT) along the interfacial direction ($x$-direction) to transform the DNS results from spatial coordinates $(x^*,y^*)$ to spectral space $(k, y^*)$, yielding the velocity field $\hat{u}(k,\, y^*, \, t^*)$ and $\hat{v}(k,\, y^*, \, t^*)$, with $k$ presenting the wave number in spectral space.
Variables marked with hat $\hat{}$ are related to DNS calculations.
Subsequently, we compute the magnitude of the system's kinetic energy in a spectral space, denoted as $\hat{E}_{\text{ke}}$, at specific time instances and $k$-th mode, by integrating the velocities in $y$-direction, 
\begin{equation}
\label{eq:int_uv}
\hat{E}_{\text{ke}}(k,t^*) = \int_{l_y} \left( |\hat{u}(k,y^*,t^*)|^{2}+|\hat{v}(k,y^*,t^*)|^{2} \right) dy^*.
\end{equation}
The expression $(k,t^*)$ represents the dependency of the variable on the $k$-th mode at time $t^*$. 
In DNS, the smallest wave number, $k_{min}$, is dependent on the length of computational domain and is defined as $k_{min}=2 \pi/(N_x \mathrm{\Delta} x)$, where $N_x$ is the number of grids in $x-$direction and $\Delta x$ is the width of grids.
The linear growth rate of kinetic energy obtained by DNS for the $k$-th mode is defined as
\begin{equation}
\label{eq:omegai_dns}
{\omega_i}(k)=\frac{1}{2}\frac{\ln \left[\hat{E}_{\text{ke}}(k, t^*+\mathrm{\Delta} t) / \hat{E}_{\text{ke}}(k,t^*)\right]}{\mathrm{\Delta} t},
\end{equation}
and the $\omega_r$ is calculated by 
\begin{equation}
\label{eq:omegar_dns}
{\omega_r}(k)=\frac{\pi}{\hat{T}(k)},
\end{equation}
where $\hat{T}(k)$ is the time period of $k$-th mode.
By applying this post-processing technique, we can calculate a series of $\omega_i$ and $\omega_r$ with different $k$ from the simulation data.

\section{Derivation of $Ma_{cr}$ and $Ca_{cr}$}
\label{appA}
\begin{figure}
    \centerline{\includegraphics[width=12cm]{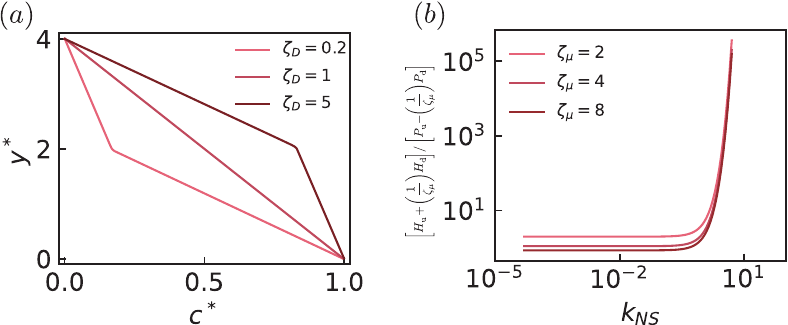}}
    \caption{ ($a$) Solute equilibrium distribution under different $\zeta_D$. 
    ($b$) The coefficient of $Ca_{cr}$ varies with $k_{NS}$ in different $\zeta_\mu$. 
    }
    \label{fig:c_cat}
\end{figure}

In equation (30) of \cite{sternling_interfacial_1959} and equation (35) of \cite{reichenbach_linear_1981}, when the $\omega_i$ is 0, the following relationship is satisfied among the dimensionless parameters, 
\begin{equation}
    \label{eq:crMa_app0}
    \left(\frac{D_u \mu_u}{\beta |G_{c}| L^2} k^2 \right)_{NS}  =\frac{\left(\zeta_D-1\right)}{8 \left(\frac{1}{\zeta_D}+1\right)\left({\frac{1}{\zeta_\mu}}+1\right)},
\end{equation}
where the superscript ${}$ means the definition in \cite{sternling_interfacial_1959} and \cite{reichenbach_linear_1981}. 
The characteristic length $L$ defined in \cite{reichenbach_linear_1981} refers to the depth of a single-layer liquid, while in this study it is half of it. 
The dimensionless result of $k$ is twice of $k$. 
In addition, $G_c$ mentioned in \cite{sternling_interfacial_1959} and \cite{reichenbach_linear_1981} is also the result of a single-sided liquid (solute input side), while in this study, the comprehensive gradient of the two side liquids is always 0.25. 
If only the solute gradient of the input side liquid is considered, a coefficient $a_{\zeta_D}$ needs to be taken into account, which is related to the $\zeta_D$, as shown in figure~\ref{fig:c_cat}.
Let the LHS becomes $Ma_{cr}$ number,
\begin{equation}
    \label{eq:crMa_app2}
    Ma_{cr} =\frac{8 k_{NS}^2\left(\frac{1}{\zeta_D}+1\right)\left({\frac{1}{\zeta_\mu}}+1\right)}{\left(\zeta_D-1\right)}  {a_{\zeta_D}}.
\end{equation}

Considering the deformable interface and limiting liquid depth, original form of eq.~(37) of \cite{reichenbach_linear_1981} is shown as following,  
\begin{equation}
    \label{eq:crMa_app3}
        M a_{cr}=\frac{\left(\epsilon^{}+\frac{1}{\zeta_D}\right)\left[H_\text{u}^{}+\left(\frac{1}{\zeta_\mu}\right)\frac{G_\text{u}^{}}{G_\text{d}^{}} H_\text{d}^{}\right]}
        {(1+\epsilon^{})\left[\frac{Cr}{ \left(k_{NS}^2+W e\right)}\right]\left[P_\text{u}^{}-\left(\frac{1 }{\zeta_\mu}\right)\frac{G_\text{u}^{}}{G_\text{d}^{}}P_\text{d}^{}\right]-\frac{1-E_\text{d}^{}}{(1-E_\text{u}^{})(1+E_\text{d}^{})} L_\text{u}^{}+\left(\frac{\zeta_D} {1+E_\text{d}^{}}\right)\frac{G_\text{u}^{}}{G_\text{d}^{}} L_\text{d}^{}},
\end{equation}
where 
$E_\text{u}=e^{-2k_{NS} l_\text{u}}$, 
$E_\text{d}=e^{-2k_{NS} l_\text{d}}$, 
$H_\text{u}=8k_{NS}^2(1 - E_\text{u}^2 - 4k_{NS}^2 l_\text{u}E_\text{u})$, 
$H_\text{d}=8k_{NS}^2(1 - E_\text{d}^2 - 4k_{NS}^2l_\text{d}E_\text{d})$, 
$G_\text{u}= -(1 - E_\text{u})^2 + 4k_{NS}^2 l_\text{u} E_\text{u}$, 
$G_\text{d}= -(1 - E_\text{d})^2 + 4k_{NS}^2 l_\text{d} E_\text{d}$, 
$L_\text{u}=(1 - E_\text{u})^3 - 4k_{NS}^3 l_\text{u}^3 E_\text{u}(1 + E_\text{u})$, 
$L_\text{d}=(1 - E_\text{d})^3 - 4k_{NS}^3 l_\text{d}^3 E_\text{d}(1 + E_\text{d})$, 
$P_\text{u}=32k_{NS}^5 l_\text{u}^2 E_\text{u}$, 
$P_\text{d}=32k_{NS}^5 l_\text{d}^2 E_\text{d}$, 
$\epsilon = [(1-E_\text{d})(1+E_\text{u})]/[(1+E_\text{d})(1-E_\text{u})]$. 
Then consider $l_\text{u}=l_\text{d}$,  when $\zeta_\mu=1$, eq.~(\ref{eq:crMa_app3}) can be simplified as,
\begin{equation}
    \label{eq:crMa_app5}
        M a_{cr}=\frac{\left(\frac{1}{\zeta_D}+1\right)}{\zeta_D-1} \frac{2 H_{\text{u}}\left({1+E_{\text{d}}}\right)}{{L_\text{d}}}a_{\zeta_D}.
\end{equation}
In the case of $\zeta_D=1$ and $\zeta_\mu \ne 1$, 
\begin{equation}
    \label{eq:crMa_app7}
    \left[-\frac{1-E_\text{d}}{(1-E_\text{u})(1+E_\text{d})} L_\text{u}+\left(\frac{\zeta_D} {1+E_\text{d}}\right)\frac{G_\text{u}}{G_\text{d}} L_\text{d}\right] = 0. 
\end{equation}
And eq.~(\ref{eq:crMa_app3}) can be simplified to,
\begin{equation}
    \label{eq:crMa_app8}
    M a_{cr}=\frac{ 2k_{NS}^2}{Cr} \frac{\left[H_\text{u}+\left(\frac{1}{\zeta_\mu}\right) H_\text{d}\right]}
    {\left[P_\text{u}-\left(\frac{1 }{\zeta_\mu}\right)P_\text{d}\right]}.
\end{equation}
Because of $Cr=Ca/Ma$, then we divide $Ma_{cr}$ in both two sides of eq.~(\ref{eq:crMa_app8}),
\begin{equation}
    \label{eq:crMa_app9}
    Ca_{cr}=2k_{NS}^2\frac{\left[H_\text{u}+\left(\frac{1}{\zeta_\mu}\right) H_\text{d}\right]}
    {\left[P_\text{u}-\left(\frac{1 }{\zeta_\mu}\right)P_\text{d}\right]}.
\end{equation}
As shown in Fig.~\ref{fig:c_cat}, the coefficient ${\left[H_\text{u}+\left(\frac{1}{\zeta_\mu}\right) H_\text{d}\right]}/
{\left[P_\text{u}-\left(\frac{1 }{\zeta_\mu}\right)P_\text{d}\right]}$ can be considered a constant when $k_{NS}<0.1$, which means the $Ca_{cr}$ is linear with ${k_{NS}^2}$. 

\section{Derivation of $\lambda_{\kappa}$}
\label{appB}
Here we perform a theoretical derivation to obtain the growth rate of $E_{\kappa}$. 
For the linearized phase field, the phase field variable $\phi_{\text{u}}$ can be decomposed as $\phi_{\text{u}}^b(y^*)+\phi_{\text{u}}^{\prime} (x^*,y^*,t^*)$, then
\begin{equation}
\label{eq:linear_phi}
\phi_{\text{u}}^{\prime} (x^*,y^*,t^*)=\tilde{\phi}_{\text{u}}(y^*)e^{ikx^*-i\omega t^*}+c.c.=e^{\omega_i t^*}\tilde{\tilde{\phi}}_{\text{u}}(x^*,y^*,t^*) .
\end{equation}
Note that $\tilde{\tilde{\phi}}_{\text{u}}$ is a function of $t^*$, but this time variation only affects the rapid oscillation, 
instead of the slow growth of the disturbance amplitude. 
From the sec.\ref{sec:2.1}, the normal vector $\boldsymbol{n}_{\text{u}}$ can be expressed as, 
\begin{equation}
\label{eq:linear_n}
\boldsymbol{n}_{\text{u}}=\frac{\nabla^* \phi_{\text{u}}}{|\nabla^* \phi_{\text{u}}|}
    =\frac{\nabla^* \phi_{\text{u}}^b+e^{\omega_i t^*}\tilde{\tilde{\phi}}_{\text{u}}}{|\nabla^* \phi_{\text{u}}^b+e^{\omega_i t^*}\tilde{\tilde{\phi}}_{\text{u}}|}
    \approx \frac{\nabla^* \phi_{\text{u}}^b+e^{\omega_i t^*}\tilde{\tilde{\phi}}_{\text{u}}}{|\nabla^* \phi_{\text{u}}^b|}
    =\frac{\nabla^* \phi_{\text{u}}^b}{|\nabla^* \phi_{\text{u}}^b|}+e^{\omega_i t^*}\frac{\tilde{\tilde{\phi}}_{\text{u}}}{|\nabla^* \phi_{\text{u}}^b|}.
\end{equation}
where we know that during the initial linear phase, $\nabla \phi_{\text{u}^b} \gg e^{\omega_i t^*} \nabla \tilde{\tilde{\phi}}_{\text{u}}$, so we can
ignore $e^{\omega_i t^*} \nabla \tilde{\tilde{\phi}}_{\text{u}}$ in the denominator. But we should not ignore that in the numerator as we are examining the initial growth.
Thus, we have
\begin{equation}
\label{eq:linear_kappa}
\kappa_{\text{u}}^* = -\nabla^* \cdot \boldsymbol{n}_{\text{u}}\approx-\nabla \cdot \left(\frac{\nabla^* \phi_{\text{u}}^b}{|\nabla^* \phi_{\text{u}}^b|}\right)  -e^{\omega_i t^*} \nabla^* \cdot \left(\frac{\tilde{\tilde{\phi}}_{\text{u}}}{|\nabla^* \phi_{\text{u}}^b|}\right)  .
\end{equation}
For present case, $-\nabla^* \cdot \left(\frac{\nabla^* \phi_{\text{u}}^b}{|\nabla^* \phi_{\text{u}}^b|}\right)=-\nabla^* \cdot \boldsymbol{e}_y=0$. 
Then,$\kappa_{\text{u}} \approx  -e^{\omega_i t^*} \nabla^* \cdot \left(\frac{\tilde{\tilde{\phi}}_{\text{u}}}{|\nabla^* \phi_{\text{u}}^b|}\right)$. 
If we take the square of it, 
\begin{equation}
\label{eq:linear_kappa_square}
\kappa_{\text{u}}^2 \approx  e^{2 \omega_i t^*} \left[\nabla^* \cdot \left(\frac{\tilde{\tilde{\phi}}_{\text{u}}}{|\nabla^* \phi_{\text{u}}^b|}\right) \right]^2 .
\end{equation}
where $e^{2\omega_i t^*}$ indicates that it will grow at the same rate as the $c^{\prime 2}$ and $\boldsymbol{u}^*\cdot \boldsymbol{u}^*$ in the linear stage because of $2\omega_i=\lambda$. 
And $\left[\nabla^* \cdot \left(\frac{\tilde{\tilde{\phi}}_{\text{u}}}{|\nabla^* \phi_{\text{u}}^b|}\right) \right]^2$ only affects the oscillation due to $\tilde{\tilde{\phi}}_{\text{u}}$. 
Through this simple derivation, we demonstrate that $\lambda$ of $E_{\kappa}$ is the same as $E_{c^\prime}$ and $E_{\text{ke}}$.

\bibliographystyle{elsarticle-num}
\bibliography{main}

\end{document}